\let\ni=\noindent

\documentclass[12 pt]{article}    
\begin{document} 
\title{ New recursion relations of matrix elements  of $r^\lambda$ and $\beta r^\lambda$ between relativistic hydrogenic eigenstates }  

\author{R. P. Mart\'{\i}nez-y-Romero\footnote{E-mail: rodolfo@dirac.fciencias.unam.mx}\\
        Facultad de Ciencias, Universidad Nacional Aut\'onoma de M\'exico,\\
        Apartado Postal 50-542, Coyoac\'an 04510, D. F., M\'exico
        \and
        H. N. N\'u\~nez-Y\'epez\footnote{E-mail: nyhn@xanum.uam.mx}\\ 
        Departamento de F\'{\i}sica,\\ 
        Universidad Aut\'onoma Metropolitana, Unidad Iztapalapa, \\
         Apar\-tado Postal  55-534, Iztapalapa 09340 D.\ F., M\'exico\\
         \and
       A. L. Salas-Brito\thanks{Corresponding author. E-mail: asb@correo.azc.uam.mx}\\
Laboratorio de Sistemas Din\'amicos,\\
Departamento de Ciencias B\'asicas,\\ Universidad
Aut\'onoma Metropolitana, Unidad Azcapotzalco,\\ Apartado
Postal 21-267,  Coyoac\'an 04000 D.\ F., M\'exico.}

\maketitle
 
\begin{abstract}
We determine exact recurrence relations which  help in the evaluation of matrix elements of 
powers of the radial coordinate between Dirac relativistic hydrogenic eigenstates. The power $\lambda$  can be any complex number as long as the corresponding term  vanishes  faster than $r^{-1}$ as $r \to \infty$. These  formulas allow  determining recursively any matrix element of radial powers ---$r^\lambda$ or $\beta r^\lambda$, $\beta$ is a Dirac matrix--- in terms of the two previous consecutive elements.  The results are useful in relativistic atomic calculations.
\end {abstract}


\section{Introduction}  \label{S:intro}

Matrix elements of the radial coordinate have been very important since the early days of quantum mechanics \cite{lowdin59,hirschfelder60,blanchard74,lange91,fernandez87,ijqc02,owono97,owono02}. These quantities are a crucial link between theoretical predictions and the observed facts. For example, they are needed for studies of the behaviour of Rydberg electrons in external fields, or in any calculation involving the multipolar expansion of electromagnetic fields \cite{vanderveldt93}.  On the other hand, the  interest in calculating relativistic effects on high Z atomic or ionic transitions\cite{grant96} requires calculating matrix elements between relativistic eigenstates of the Dirac hydrogen atom.   Experiments with precisions of the order of  $ 10^{-3}$ eV or less are now standard. For example,  experiments performed using new photon sources for the Opacity and Iron Projects, or experiments with merged beam techniques, or the use of specialized laser sources \cite{aguilar02,west97,mueller02}. All of these mean that relativistic effects can be quite easily observed in atomic, molecular or ionic processes and, therefore, that  techniques for evaluating relativistic expectation values are quite useful---as has been known for some time now \cite{oumarou95}.

The nonrelativistic evaluation of   matrix elements has a long and succesful history, but the relativistic efforts have mainly come from the last 35 years or so. We have to pinpoint however, that even these relativistic calculations, even exact ones, are an approximation, for the only way of calculating exactly relativistic effects in the interaction of atoms with electromagnetic fields is using the full QED formalism. See, for example, \cite{spruch94}. But even so, many calculations use a  nonrelativistic approach recurring to the Schr\"odinger equation ---which is also a valid approximation  \cite{chen03}. Even in the relativistic quantum mechanics approximation, in many circunstances it is much better to have recursion relations between power-of-$r$ terms than to have to deal with the cumbersome and complex formulas that stem from the exact evaluation of the matrix elements \cite{jpb01}. Furthermore, all what is needed in many ocassions is to have the matrix elements for exponents in a certain range, so it is very convenient to have recursion formulas \cite{jpb02}. For instance, the behaviour of Rydberg electrons in external fields, or the long range interactions of ionic cores \cite{cores}. Thus, we have been investigating such formulas in Dirac relativistic quantum mechanics. We have discovered various of such expressions since, perhaps not surprisingly, there appears to be more independent recursions. We have been applying an hypervirial inspired technique that have produced excelent results in nonrelativistic quantum mechanics \cite{jpb95} to uncover some of the recurrence relations, but we are now combining that technique with  operator algebra \cite{fernandezcastro00} to generalize the previously obtained recurrence relations to more compact  recursions; equations (\ref{74}) and (\ref{73}) in section \ref{S:rr}. An advantage of these recursions is that they are able to relate only three consecutive powers of $r$ or $\beta r$---to be compared with the results in \cite{ijqc02,jpb02}

Our paper is organized as follows. In section \ref{S:ha} the relativistic quantum mechanics of an hydrogen atom is reviewed since all the discussions that follows use its energy eigenfunctions and eigenvalues. A useful feature of our discussion is that we express compactly  the relativistic eigenfunctions in terms of generalized Laguerre polynomials of noninteger index \cite{jmp99,davies39}. The relativistic recurrence relations of matrix elements of powers of the radial coordinate are derived in section \ref{S:rr}. Section \ref{S:coeff} gathers the definitions of the symbols we use for writing compactly these recurrence relations.  Section \ref{S:conc} contains our conclusions. 
 
\section{The relativistic hydrogen atom}  \label{S:ha}
     
The radial wave function of an electron in an hydrogen atom  is

\begin{equation}
\Psi= \frac{1} {r} \pmatrix{F_{nj\epsilon}(r)\cr i G_{nj\epsilon}(r)}
\end{equation}

\ni where  $F_{nj\epsilon}(r)$, $G_{nj\epsilon}(r)$ are respectively called the big and the small components of the spinorial wavefunctions $\Psi$, $n=1,2,3,\dots$, $j=1/2,3/2,5/2,\dots$ is the total (orbital plus spin) angular momentum quantum number of the electron, and $\epsilon=(-1)^{j+l-1/2}$, $l=j\pm1$ is the orbital angular momentum quantum number, the sign is chosen according to whether $l$ refers to the big ($+$) or to the small ($-$) component. This wave function is a solution of the Coulomb radial Dirac equation

\begin{equation}
\left(c\alpha_r\left[p_r-i\frac{\hbar\beta}{r}\epsilon\left(j+\frac{1}{2}\right)\right]+\beta m c^2 -\frac{Z q_e^2}{r}\right)\Psi(r)=E\Psi(r)
\end{equation}

\ni where $Z$ is the atomic number, $m$ the electron mass, $q_e$ is the magnitude of the electron charge, and the subscript $r$ refers to the radial part of the wave function. The Dirac matrices are

\begin{equation}
\alpha_r=\pmatrix{0& -1\cr -1& 0},\qquad \beta=\pmatrix{1& 0\cr 0& -1},
\end{equation}

\ni and the radial momentum operator is 

\begin{equation}
p_r=-i\hbar \left(\frac{1}{r}+\frac{d}{dr}\right).
\end{equation}

The components of the eigenfunctions, $F$ and $G$, are the solutions of the following differential equations \cite{jmp99,ajp00,jpa98}

\begin{equation}
\left(-\frac{d}{d\rho} + \frac{\epsilon(j+1/2)}{\rho} \right)G(\rho)=\left(-\nu+\frac{Z\alpha_F} {\rho} \right) F(\rho)
\end{equation}

\ni and

\begin{equation}
\left(+\frac{d}{d\rho} + \frac{\epsilon(j+1/2)}{\rho} \right)F(\rho)=\left(\frac{1}{\nu}+\frac{Z\alpha_F} {\rho} \right) G(\rho)
\end{equation}

\ni where 
\begin{equation}
\rho=kr,\qquad
k\equiv \frac{1}{\hbar c}\sqrt{m^2c^4-E^2},\qquad \nu \equiv \sqrt{\frac{mc^2-E}{mc^2+E}}.
\end{equation}

\ni The solutions of these coupled equations ---that can be expressed in terms of the Sonine polynomials, in terms of Laguerre polynomials, of non-integer index, or as special cases of the hypergeometric function, see \cite{jmp99,davies39,ajp00,jpa98,draganescu02} for details---  can be written as

\begin{equation}
F(\rho)=+C \sqrt{mc^2+E} \rho^s \exp(-\rho)\left[A_n L^{2s}_{n}(2\rho)+B_{n-1}^{2s}(2\rho)\right],
\end{equation}
\ni and
\begin{equation}
G(\rho)= -C \sqrt{mc^2-E} \rho^s \exp(-\rho)\left[A_n L^{2s}_{n}(2\rho)-B_{n-1}^{2s}(2\rho)\right],
\end{equation}

\ni where the $L_n^{2s}(x)$ are Laguerre polynomials of noninteger index, the numbers $A_n$ and $B_n$ are
\begin{equation}
A_n=\left[\epsilon\left(j+\frac{1}{2}\right)+s+\frac{Z\alpha_F}{\nu}+n-\frac{Z\alpha_F}{\nu}\right]^{1/2},
\end{equation}
\begin{equation}
B_n=(n+2s)\left[\epsilon\left(j+\frac{1}{2}\right)+s-\frac{Z\alpha_F}{\nu}+n-\frac{Z\alpha_F}{\nu}\right]^{1/2};
\end{equation}

\ni where $s=\sqrt{(j+1/2)^2-Z^2\alpha_F^2}$.  The normalization constant $C$ is

\begin{equation}
C=\frac{\hbar 2^{s-1}}{Z\alpha_{{}_F}c^2}\sqrt{\frac{n!k}{2m^3\Gamma(n+2s+1)}}.
\end{equation}

The energy levels are given by the usual expression \cite{ajp00,bethe57} 

\begin{equation}\label{energyeigenvalues}
\frac{E}{mc^2}\equiv \frac{E_{nj\epsilon}}{mc^2}=\left[1+\frac{Z^2\alpha_F^2}{n-j-1/2+\sqrt{(j+1/2)^2-Z^2\alpha_F^2}}  \right]^{-1/2},
\end{equation}

\ni where $n=1,2,3,\dots$, $\alpha_F= q_e^2/\hbar c\simeq 1/137$ is the fine structure constant. It is convenient to point out that the quantum number $\epsilon$ we use is related to the often used $\kappa$ \cite{grant96} as $\kappa=-\epsilon(j+1/2)$. The quantum number $\kappa$ is an eigenvalue of the following operator 

\begin{equation}
\hat{K}=\beta\left(\hbar^2+\mathbf{\Sigma}\cdot{\bf L}\right), \quad \hbox{where}\quad {\mathbf \Sigma}= 2{\mathbf S}= \hbar\pmatrix{{\mathbf \sigma}& 0\cr 0& {\mathbf\sigma}},
\end{equation}

\ni and ${\mathbf \sigma}=(\sigma_x, \sigma_y, \sigma_z)$ is the standard 3-vector of the Pauli matrices. For the sake of simplicity we often use the single symbol $a$ to mean all the three quantum numbers $n_a, j_a,\epsilon_a$, and write the radial wave functions in ket (or bra) form as $\Psi_a\equiv |a\rangle$.

We calculate recursions beween matrix elements of the form

\begin{equation}
\langle a'|\beta^b r^\lambda| a\rangle=\int dr\; r^2 \Psi_{a'}^{\dagger}(r) \beta^b r^\lambda \Psi_a(r)
\end{equation}

\ni where $\Psi_{a'}^{\dagger}\equiv (F_{a'}^*/r, -iG_{a'}^*/r)$,   $b=0,\hbox{ or }1$, and $\lambda$ is a possibly complex exponent. The recurrence relations (\ref{III}), (\ref{beta}), (\ref{70}) --(\ref{73}) hold true if $s_1 +s_2 +1 +|\lambda| >0$ \cite{jpb01,jpb00} where as before the $s_b\equiv+\sqrt{(j_b+1/2)^2-Z^2\alpha^2_F},\; b=1,2$ are real numbers; this conditions is basically  that any  integrand goes to zero faster than $1/r$ as $r\to\infty$ \cite{jpb02}.

One of the advantages of the approach developed here is that is not only applicable to Dirac hydrogenic wavefunctions but can  also be used within  the quantum defect approximation, as discussed by Owono Owono, Kwato Njock, and Oumarou \cite{owono02}, see also \cite{owono97}. This feature is important since the quantum deffect approximation describes accurately the behaviour of Rydberg electrons in atomic systems \cite{owono97,owono02,nieto95,nieto952,karwowski91}. 

\section{Relativistic recursion relations}\label{S:rr}

Using hypervirial methods we have recently obtained a useful form of recurrence relations for atomic eigenstates. In this work we want to discuss the additional use of operator algebra to obtain alternative and independent expressions for recurrence relations of atomic expectation values. Let us begin with the radial Dirac equation for an electron in an arbitarry  radial potential $V(r)$, where the index $a$ serves to distinguish Hamiltonians with different quantum numbers and it is convenient to distinguish wavefunctions with different energy eigenvalues. 

Let us begin with the Dirac  radial equation written with a certain radial potential that afterwards will be identified with Coulomb's

\begin{equation}
H_a\Psi_a=\left(c\alpha_r\left[p_r-i\frac{\hbar \beta}{r}\epsilon_a\left(j_a+\frac{1}{2}\right)\right]+\beta m_a c^2 + V_a(r)\right)\Psi_a(r)=E_a\Psi_a(r),
\end{equation}

\ni here $H_a$ is the Dirac Hamiltonian,  the energy eigenvalue $E_a$ is given by (\ref{energyeigenvalues}), and $a$ is an index that will be quite useful in what follows. 

 To begin with, let us  consider an arbitrary radial function $f(r)$ and define $\xi(r)\equiv H_2f-fH_1$. To obtain the recurrence relations, let us  establish that

\begin{eqnarray}\label{3} 
H_2 \xi &-& \xi H_1 = -c^2 \hbar^2f''-c^2\hbar^2\beta \frac{\Delta^-}{2r} \left(2f\frac{d}{dr}+f'+\frac{f}{r}\right) + c^2\hbar^2\beta \frac{\Delta^+}{2r} f' \nonumber\\
&+&c^2\hbar^2 \left(\frac{\Delta^-}{2r}\right)^2 f + \left(c^2 \beta m^- +V^-\right)^2 f \nonumber\\
&-&i \hbar c \alpha_r \left[ \left(f'+\frac{\Delta^-}{2r}\beta f \right) \left(V^- -c^2\beta m^+\right) + V^- f' +(V^-)'f \right. \nonumber\\
&+&\left.c^2\beta m^- \left(2f'\frac{d}{dr} +f'+2\frac{f}{r}\right) +c^2m^-\frac{\Delta^+}{2r}f + \beta V^-\frac{\Delta^-}{2r}f \right],
\end{eqnarray}

\ni where, $\Delta_a\equiv \epsilon_a(2j_a+1)$ and, if $X$ is any symbol of interest, we are ---and will be--- using  

\begin{equation}
X^{\pm}\equiv X_2\pm X_1.  
\end{equation}

The following five identities are not too difficult to establish by the use of a little operator algebra

\begin{eqnarray}\label{6}
H_2(-i c\alpha_r f) &+& (-i c\alpha_r f)H_1 = 
- c^2\hbar \left(2f\frac{d}{dr}+f'+2\frac{f}{r} -\frac{\Delta^-}{2r}\beta f \right) \nonumber\\
&+& i c\alpha_r \left(c^2 \beta m^- -V^+\right) f,
\end{eqnarray}
\begin{equation}\label{1}
H_2f - fH_1 = -i \hbar c \alpha_r \left(f'+\frac{\Delta^-}{2r}\beta f \right) 
+\left(c^2 \beta m^- + V^-\right) f,
\end{equation}
\begin{equation}\label{2}
H_2f + fH_1 = -i \hbar c \alpha_r \left(2f\frac{d}{dr} + f'+ 2\frac{f}{r}
+\frac{\Delta^+}{2r}\beta f \right) +\left(c^2 \beta m^+ + V^+\right) f,
\end{equation}
\begin{equation}\label{13}
H_2V^-f - V^-fH_1 = -i \hbar c \alpha_r \left(V^-f'+(V^-)'f+\frac{\Delta^-}{2r}V^-\beta f \right) 
+\left(c^2 \beta m^- + V^-\right) V^-f,
\end{equation}
\ni and 

\begin{eqnarray}\label{12}
H_2\left(-i c\alpha_r \beta \frac{f}{r}\right) &+& \left(-i c\alpha_r \beta \frac{f}{r}\right)H_1 = 
- c^2\hbar \left[\beta \left(\frac{f'}{r}-\frac{f}{r^2}\right) -\frac{\Delta^+}{2r}\frac{f}{r} \right] \nonumber\\
&+& i c\alpha_r \left(c^2 \beta m^- -V^+\right)\beta \frac{f}{r}.
\end{eqnarray}

  To get the relation we are after, we need to eliminate the terms involving $i c \alpha_r$ and the terms involving $d/dr$ of equation (\ref{3}). To this end, from (\ref{6}) we extract the term in the left hand side (LHS) of the following equation

\begin{eqnarray}\label{24}
&-& c^2\hbar^2 \beta \frac{\Delta^-}{2r} \left(2f\frac{d}{dr}+f'+2\frac{f}{r}\right) = - i \hbar c \alpha_r \beta \frac{\Delta^-}{2r} \left( c^2 \beta m^- -V^+\right) f \nonumber\\
&-& c^2 \hbar^2 \left( \frac{\Delta^-}{2r} \right)^2 f - \left(H_2(-i \hbar c\alpha_r \beta \frac{\Delta^-}{2r} f) + (-i\hbar c\alpha_r \beta \frac{\Delta^-}{2r}f)H_1 \right).
\end{eqnarray}

\ni Furthermore, from (\ref{1}), we extract 

\begin{equation}\label{1d}
-i \hbar c \alpha_r \left(f'+\frac{\Delta^-}{2r}\beta f \right) = 
- \left(H_2f - fH_1\right) - \left(c^2 \beta m^- + V^-\right) f, 
\end{equation}

\ni from (\ref{2}), we extract

\begin{eqnarray}\label{21}
&-&i \hbar  \alpha_r c^3 \beta m^- \left(2f\frac{d}{dr} + f'+ 2\frac{f}{r}\right) = 
- \left(H_2 c^2 m^- \beta f + c^2 m^- \beta fH_1 \right) \nonumber\\
&+&i \hbar c \alpha_r c^2 m^- \frac{\Delta^+}{2r} f -\left(c^2 \beta m^+ + V^+\right)\beta f, 
\end{eqnarray}

\ni from (\ref{13}), we extract

\begin{eqnarray}\label{13d}
&-& i \hbar c \alpha_r \left(V^-f'+(V^-)'f\right) =  \left( H_2V^-f - V^-fH_1 \right) \nonumber\\
&+&i \hbar c \alpha_r \frac{\Delta^-}{2r}V^-\beta f  
-\left(c^2 \beta m^- + V^-\right) V^-f,
\end{eqnarray}

\ni and from (\ref{12}), we extract

\begin{eqnarray}\label{29}
 &&\left[H_2\left(-i c\hbar \alpha_r \frac{\Delta^-}{2r} \beta \frac{f}{r}\right) + \left(-i c \hbar \alpha_r \frac{\Delta^-}{2r} \beta \frac{f}{r}\right)H_1\right] - \nonumber
\\  && i c\hbar \alpha_r \frac{\Delta^-}{2r} \left(c^2 \beta m^- -V^+\right)\beta f 
=  - c^2\hbar^2 \frac{\Delta^-}{2r} \left[\beta \left(f'- \frac{f}{r} \right) - \frac{\Delta^+}{2r}f \right]. 
\end{eqnarray}

\ni The  LHS of equations (\ref{24}) to (\ref{29})  are to be substituted into equation (\ref{3}). Then, after introducing explicitly the Coulomb potential $V_1=V_2=-Zq_e^2/r$, imposing the equality of the masses $m_1=m_2$ (since $H_1$ and $H_2$ describe the same system),  substituting the arbitrary function $f(r)$ for the potential function $r^\lambda$, and  taking matrix elements between states $\langle2|$ and $|1\rangle$ at the end, these steps  yield the recurrence relation

\begin{equation}\label{III}
\left(E_2-E_1\right)^2 \langle2|r^\lambda |1\rangle+ k_2 \langle2|r^{\lambda-2}|1\rangle = l_0 \langle2|\beta r^\lambda |1\rangle+ l_2 \langle2|\beta r^{\lambda-2}|1\rangle,
\end{equation} 

\ni where 

\begin{eqnarray}\label{IIIc} 
k_2 &=& c^2 \hbar^2 \lambda \left( \lambda - 1 \right) - \frac{c^2 \hbar^2}{4} \Delta^+ \Delta^-, \nonumber\\ 
l_0 &=& -2 c^2 m (E_2-E_1), \nonumber\\ 
l_2 &=& \frac {c^2 \hbar^2}{2} \left[ 2 \Delta^- + \lambda \left( \Delta^+ - \Delta^- \right) \right].
\end{eqnarray}

  Now, following a strictly similar procedure but substituting equations (\ref{6}), (\ref{1}), (\ref{12}), and the following one

\begin{eqnarray}\label{14} 
&& H_2 V^- f + V^- f H_1 = \left(c^2 \beta m^+ + V^+\right) V^- f \nonumber\\
&-& i \hbar c \alpha_r \left[2 V^- f \frac{d}{dr} + V^-f'+(V^-)'f + 2 V^- \frac{f}{r} + V^- \frac{\Delta^+}{2r}\beta f \right], 
\end{eqnarray}

\ni into the next equation 

\begin{eqnarray}\label{4} 
H_2 \xi &+& \xi H_1 = -c^2 \hbar^2 \left(2f'\frac{d}{dr}+f''+2\frac{f'}{r}-\beta \frac{\Delta^-}{2r} \frac{f}{r}\right) \nonumber\\
&+& c^2 \hbar^2 \frac{\Delta^+}{2r} \frac{\Delta^-}{2r}f + \left( c^2 \beta m^+ + V^+ \right) \left( c^2 \beta m^- + V^- \right)f  
 \nonumber\\
&-&i \hbar c \alpha_r \left[\left(f'+\frac{\Delta^-}{2r}\beta f \right) \left(V^+ -c^2\beta m^- \right) + c^2 \beta m^-f' + c^2m^- \frac{\Delta^-}{2r} f \right. \nonumber\\
&+& \left. 2 V^- f \frac{d}{dr} + V^-f'+(V^-)'f + 2 V^- \frac{f}{r} + V^- \frac{\Delta^+}{2r}\beta f \right],
\end{eqnarray}

\ni  again introducing explicitly the Coulomb potential $V_1=V_2=-Zq_e^2/r$, making the masses equal $m_1=m_2$,  substituting  the potential function $r^\lambda$, and, at the end, taking the  matrix elements, we  obtain the following recurrence relation  

\begin{equation}\label{beta}
\left(E_2+E_1\right) \left(E_2-E_1\right)\langle2|r^\lambda|1\rangle + M_1 \langle2|r^{\lambda-1}|1\rangle + M_2 \langle2|r^{\lambda-2}|1\rangle = N_2 \langle2|\beta r^{\lambda-2}|1\rangle,
\end{equation}

\ni where 

\begin{eqnarray}\label{betac} 
M_1 &=& 2 E^- Z q_E^2, \nonumber\\ 
M_2 &=& c^2 \hbar^2 \lambda \left( \lambda - 2 \right) - \frac{c^2 \hbar^2}{4} \Delta^+ \Delta^-, \nonumber\\ 
N_2 &=& \frac{c^2 \hbar^2}{2} \left[ \Delta^- + \lambda \left( \Delta^+ - \Delta^- \right) \right].
\end{eqnarray}

\ni The new recursions [equations (\ref{III}) and (\ref{beta})] can be more useful if the matrix elements of terms $r^\lambda$ were uncoupled from the matrix elements of terms $\beta r^{\lambda'}$. 
To disentangle such relations, we need the following three previously reported recursions
\cite{ijqc02,jpb02},

\begin{equation}\label{I}
c_0 \langle2|r^\lambda|1\rangle = c_1 \langle2|r^{\lambda-1}|1\rangle + c_2 \langle2|r^{\lambda-2}|1\rangle + c_3 \langle2|r^{\lambda-3}|1\rangle = d_2 \langle2|\beta r^{\lambda-2}|1\rangle + d_3 \langle2|\beta r^{\lambda-3}|1\rangle,
\end{equation}

\begin{equation}\label{II}
e_0 \langle2|\beta r^\lambda|1\rangle = b_0 \langle2|r^\lambda|1\rangle + b_2 \langle2|r^{\lambda-2}|1\rangle + e_1 \langle2|\beta r^{\lambda-1}|1\rangle + e_2 \langle2|\beta r^{\lambda-2}|1\rangle,
\end{equation}

\ni and 

\begin{equation}\label{alfa}
g_2 \langle2|r^{\lambda-2}|1\rangle = p_2 \langle2|\beta r^{\lambda-2}|1\rangle + p_3 \langle2|\beta r^{\lambda-3}|1\rangle.
\end{equation}

\ni where the coefficients $b$, $c$,  $d$, $e$, $g$,  and $p$ are all defined in \cite{jpb02}.

 Using now (\ref{beta}) and (\ref{alfa}) in (\ref{III}), (\ref{I}) and (\ref{II}), 
we get the following six uncoupled recurrence relations,

\begin{equation}\label{70}
{\tt A_0} \langle2|r^\lambda|1\rangle = {\tt A_1} \langle2|r^{\lambda-1}|1\rangle + {\tt A_2} \langle2|r^{\lambda-2}|1\rangle + {\tt A_3} \langle2|r^{\lambda-3}|1\rangle, 
\end{equation}

\begin{equation}\label{71}
{\tt B_0} \langle2|r^\lambda|1\rangle = {\tt B_1} \langle2|r^{\lambda-1}|1\rangle + {\tt B_2} \langle2|r^{\lambda-2}|1\rangle + {\tt B_3} \langle2|r^{\lambda-3}|1\rangle. 
\end{equation}

\begin{equation}\label{72}
{\tt C_0} \langle2|r^\lambda|1\rangle = {\tt C_1} \langle2|r^{\lambda-1}|1\rangle + {\tt C_2} \langle2|r^{\lambda-2}|1\rangle + {\tt C_3} \langle2|r^{\lambda-3}|1\rangle + {\tt C_4} \langle2|r^{\lambda-4}|1\rangle,
\end{equation}

\begin{equation}\label{67}
{\tt D_0} \langle2|\beta r^\lambda|1\rangle = {\tt D_1} \langle2|\beta r^{\lambda-1}|1\rangle + {\tt D_2} \langle2|\beta r^{\lambda-2}|1\rangle + {\tt D_3} \langle2|\beta r^{\lambda-3}|1\rangle + {\tt D_4} \langle2|\beta r^{\lambda-4}|1\rangle, 
\end{equation}

\begin{equation}\label{68}
{\tt E_0} \langle2|\beta r^\lambda|1\rangle = {\tt E_1} \langle2|\beta r^{\lambda-1}|1\rangle + {\tt E_2} \langle2|\beta r^{\lambda-2}|1\rangle + {\tt E_3} \langle2|\beta r^{\lambda-3}|1\rangle, 
\end{equation}

\ni and 

\begin{equation}\label{69}
{\tt F_0} \langle2|\beta r^\lambda|1\rangle = {\tt F_1} \langle2|\beta r^{\lambda-1}|1\rangle + {\tt F_2} \langle2|\beta r^{\lambda-2}|1\rangle + {\tt F_3} \langle2|\beta r^{\lambda-3}|1\rangle. 
\end{equation}

\ni The equation obtained from substituting (\ref{alfa}) into (\ref{II}) [equation (\ref{68}) above] coincides with one of the previously obtained recursions [equation (18) in Ref.\ \cite{jpb02}]. This is nice since it serves as a way of checking the above recursions. The coefficients in equations (\ref{70}), (\ref{71}), (\ref{72}), (\ref{67}), (\ref{68}), and (\ref{69}) are all gathered in section \ref{S:coeff}.

The equations  (\ref{70}) and (\ref{71})  above  express  two forms of a recurrence relation relating matrix elements of $r^\lambda$ with matrix elements of $r^{\lambda-i}$, $i=1,2,3$. This is  a manifestation of the richer behaviour  --- though at the same time more restricted --- of relativistic quantum mechanics as compared with the nonrelativistic theory. We can combine anyway   equation (\ref{70}) with equation (\ref{71}) to get the simpler relation

\begin{equation}\label{74}
\left( \frac{{\tt B}_0}{{\tt B}_3}- \frac{{\tt A}_0}{{\tt A}_3}\right) \langle2|r^\lambda|1\rangle=\left( \frac{{\tt B}_1}{{\tt B}_3}- \frac{{\tt A}_1}{{\tt A}_3}\right)\langle2|r^{\lambda-1}|1\rangle + \left( \frac{{\tt B}_2}{{\tt B}_3}- \frac{{\tt A}_2}{{\tt A}_3}\right)\langle2|r^{\lambda-2}|1\rangle.
\end{equation}

 \ni   A   similar situation occurs with equations (\ref{68}) and (\ref{69}) above; they can be  combined to yield the simpler relation

\begin{equation}\label{73}
\left( \frac{{\tt F}_0}{{\tt F}_3}- \frac{{\tt E}_0}{{\tt E}_3}\right) \langle2|\beta r^\lambda |1\rangle=\left( \frac{{\tt F}_1}{{\tt F}_3}- \frac{{\tt E}_1}{{\tt E}_3}\right)\langle2|\beta r^{\lambda-1} |1\rangle + \left( \frac{{\tt F}_2}{{\tt F}_3}- \frac{{\tt E}_2}{{\tt E}_3}\right)\langle2|\beta r^{\lambda-2}|1\rangle.
\end{equation}

\ni Equations (\ref{73}) and (\ref{74}) are simpler, and thus potentially more useful,  than those reported in \cite{ijqc02,jpb02}.

\section{The coefficients in the recursion relations}\label{S:coeff}

The coeficients used in the recursion relations [equations (\ref{70}), (\ref{71}), (\ref{72}), (\ref{67}), (\ref{68}), and (\ref{69})] are defined as ---the various symbols used to write these equations are defined at the end in equation (\ref{def}).

\begin{eqnarray}\label{70c}
{\tt A_0} &=& \frac{\hbar E^+ \left( E^- \right)^2 \Delta^-}{D} - \frac{\hbar E^+E^-\Delta^-W}{R}, \nonumber\\
{\tt A_1} &=& -\frac{2E^-K}{D+4\hbar mc^2} + W \frac{2K}{R} + \frac{(\lambda-1) K}{D+4\hbar mc^2} \left[ \frac{2 \hbar E^+ \Delta^+}{R-\hbar (\Delta^+ - \Delta^-)} \right], \nonumber\\
{\tt A_2} &=& \frac{c^2 \hbar \Delta^- P}{2} + c^2 \hbar W \frac{\Delta^- S}{4 R} + \frac{(\lambda-1)K}{D+4\hbar mc^2} \left[ \frac{4\hbar Z q_e^2 \Delta^+}{R-\hbar (\Delta^+ - \Delta^-)} \right], \nonumber\\
{\tt A_3} &=& \frac{c^2 \hbar \Delta^- P}{2} + \frac{Q}{D+4\hbar mc^2} \left[\frac{S+12\hbar^2 (1-\lambda)}{2\hbar^2 \left[ \Delta^+ - \lambda (\Delta^+ - \Delta^-)\right] }\right],
\end{eqnarray}

\ni The $\tt A_a$ are the explicit coefficients in equation (\ref{70}).\par

\begin{eqnarray}\label{71c}
{\tt B_0} &=& -8Zq_e^2 D \frac{E^+ E^-}{c^2 \hbar R}, \nonumber\\
{\tt B_1} &=& 4\hbar \lambda F -\frac{16(Zq_e^2)^2 E^- D}{c^2 \hbar^2 [\Delta^+ + \lambda(\Delta^+ + \Delta^-)]} - \frac{\Delta^+ E^+ E^- L}{R} - \frac{4(E^+)^2 E^- D}{c^2 \hbar [R+2 \hbar (\Delta^+ - \Delta^-)]}, \nonumber\\
{\tt B_2} &=& -2 A q_e^2 D \frac{4(\lambda^2-1) - \Delta^+ \Delta^-}{[\Delta^+ + \lambda (\Delta^+ - \Delta^-)]} - \frac{2 Zq_e^2 E^- \Delta^+ L}{R} - \frac{2 Zq_e^2 E^- E^+ D}{c^2\hbar[R+2\hbar (\Delta^+ - \Delta^-}, \nonumber\\
{\tt B_3} &=& -c^2 \hbar (1-\lambda)L - \frac{c^2}{4 R} \Delta^+ L \left(S-4\hbar \lambda \right) -2 E^+ D \frac{S+12\hbar^2 \lambda}{2\hbar^2 [R+2\hbar (\Delta^+ - \Delta^-)]},
\end{eqnarray}

\ni The $\tt B_a$ are the explicit coefficients in equation (\ref{71}).\par

\begin{eqnarray}\label{72c}
{\tt C_0} &=& - \frac{4m E^+ \left( E^- \right)^2}{\hbar[R+2\hbar (\Delta^+ - \Delta^-)]}, \nonumber\\
{\tt C_1} &=& - \frac{8m Zq_e^2 (E^-)^2}{\hbar[R+2\hbar (\Delta^+ - \Delta^-)]}, \nonumber\\
{\tt C_2} &=& -mc^2 E^- \frac{S+12\hbar^2 \lambda}{\hbar[R+2\hbar (\Delta^+ - \Delta^-)]}, \nonumber\\
{\tt C_3} &=& 2Zq_e^2 E^- \frac{R + \hbar \Delta^-}{R}, \nonumber\\
{\tt C_4} &=& \frac{c^2}{4 R} \left(S-4\hbar^2 \lambda \right) \left(R + \hbar \Delta^- \right) - c^2 \hbar^2 \lambda (\lambda - 1) + \frac{c^2 \hbar^2}{4}\Delta^+ \Delta^-.
\end{eqnarray}

\ni The $\tt C_a$ are the explicit coefficients in equation (\ref{72}).\par

\begin{eqnarray}\label{67c}
{\tt D_0} &=& \frac{2}{D} \left(\frac{\Delta^-}{\Delta^+}\right) E^+ \left( E^- \right)^2 \left(T+\hbar \right), \nonumber\\ 
{\tt D_1} &=& - \frac{4Zq_e^2 \left(E^-\right)^2}{D} \left(\frac{\Delta^-}{\Delta^+}\right) T - \frac{Zq_e^2 \hbar \Delta^-}{D} E^+ \left( E^- \right)^2 U, \nonumber\\ 
{\tt D_2} &=& - \frac{2 \hbar}{D} \left(E^-\right)^2 \left(Z_e^2\right)^2 \Delta^- U + \left(\frac{\Delta^-}{\Delta^+}\right) c^2 \left(T+\hbar \right) P + \frac{c^2\hbar^2}{2} \Delta^- W, \nonumber\\ 
{\tt D_3} &=& \frac{c^2 \hbar}{2} Zq_e^2 \Delta^- U P - \frac{2}{\hbar \Delta^+ D} \left(T-2\hbar \right) Y + \frac{Q}{D+4\hbar mc^2}, \nonumber\\ 
{\tt D_4} &=& - \frac{Zq_e^2 Y E^+ \left( E^- \right)^2}{D+4\hbar mc^2}, 
\end{eqnarray}

\ni The $\tt D_a$ are the explicit coefficients in equation (\ref{67}).\par

\begin{eqnarray}\label{68c}
{\tt E_0} &=& 2 E^+ D - \frac{8 \lambda}{\Delta^+} F \left( T + \hbar \right), \nonumber\\ 
{\tt E_1} &=& - 4 Z q_e^2 D + \frac{4 \hbar}{\Delta^+} Z q_e^2 \lambda F U, \nonumber\\ 
{\tt E_2} &=& - \frac{c^2 \hbar}{2} \Delta^+ L - \frac{2 c^2}{\Delta^+} L \left(1 - \lambda \right) \left( T - \hbar \right), \nonumber\\ 
{\tt E_3} &=& - \frac{c^2 Z q_e^2}{\Delta^+} \left( 1 - \lambda \right) L U,
\end{eqnarray} 

\ni The $\tt E_a$ are the explicit coefficients in equation (\ref{68}).\par

\begin{eqnarray}\label{69c}
{\tt F_0} &=& \frac{4 m c^2}{\hbar \Delta^+} \left(E^-\right)^3 \left(T+\hbar \right), \nonumber\\ 
{\tt F_1} &=& Zq_e^2 \left(E^-\right)^2 U, \nonumber\\ 
{\tt F_2} &=& \frac{c^2 E^+}{8 \hbar \Delta^+} S U + \frac{c^2}{2 \Delta^+} S \left(\lambda-1 \right)+ \frac{c^2 \hbar}{2} \left(R + \hbar \Delta^-\right), \nonumber\\ 
{\tt F_3} &=& \frac{c^2}{4} Zq_e^2 S U,
\end{eqnarray} 

\ni The $\tt F_a$ are the explicit coefficients in equation (\ref{69}).\par

 The above equations are written in terms of the following symbols.

\begin{eqnarray}\label{def}
D &=& \hbar \Delta^- E^- - 4mc^2 \hbar \lambda, \nonumber\\ 
T &=& \frac{\Delta^+ + \Delta^-}{4mc^2} \hbar E^+ + \hbar \lambda, \nonumber\\ 
F &=& \left( E^- \right)^2 -4 m^2 c^4, \nonumber\\ 
K &=& Zq_e^2 \hbar E^- \Delta^-, \nonumber\\ 
L &=& 4 \hbar^2 \lambda^2 - \hbar^2 \left( \Delta^- \right)^2, \nonumber\\ 
U &=& \frac{\hbar}{m c^2} \left(\Delta^+ + \Delta^- \right), \nonumber\\ 
W &=& \left( 1 - \lambda \right) + \frac{\lambda E^+ \Delta^+}{\Delta^- E^- - 4mc^2 \lambda}, \nonumber\\ 
Y &=& 2 c^2 \hbar^3 Z q_e^2 \left( \lambda - 1 \right) \left( \lambda - 2 \right) \Delta^-, \nonumber\\ 
P &=& \frac{Y}{D} - \frac{\hbar \Delta^+}{2}, \nonumber\\ 
Q &=& c^2 \hbar^3 Z q_e^2 \left( \lambda - 1 \right) \Delta^- \Delta^+, \nonumber\\ 
S &=& 4 \hbar^2 \lambda \left( \lambda - 1 \right) - \hbar^2 \Delta^+ \Delta^-, \nonumber\\ 
R &=& \hbar \Delta^- + \lambda \hbar \left( \Delta^+ - \Delta^- \right).
\end{eqnarray}

The coefficients appearing in equations (\ref{74}) and (\ref{73}) are respectively expressed in terms of definitions (\ref{70c}), (\ref{71c}), and  (\ref{68c}),  (\ref{69c}).

\section{Conclusion}\label{S:conc} In this paper we have derived exact recurrence relations between general non-necessarily diagonal  matrix elements of powers of the radial coordinate. These recursions  relate any three consecutive powers of $r$ or of $\beta r$. The states used for evaluating the matrix elements are radial completely relativistic hydrogenic eigenstates. The derivation was done employing a technique inspired in the hypervirial method and using some operator algebra \cite{hirschfelder60,blanchard74,lange91,jpb95}. The matrix elements analysed here are all gauge invariant \cite{owono02}. The relations obtained may have different uses in many interesting atomic calculations, as in the calculation of transitions between Rydberg states beyond the semiclassical approximation \cite{oumarou95,oumarou91}. 
\section*{Acknowledgements}

This work has been partially supported by PAPIIT-UNAM (grant 108302). We acknowledge with thanks the insightful comments  of P T M Jarel, G R Inti, G R Maya, A S Ubo, and P M Schwartz.

\end{document}